\documentclass[referee,a4paper,12pt,traditabstract]{jswsc} 

\usepackage{graphicx}
\usepackage{txfonts}
\usepackage{subfigure}
\usepackage{epstopdf}
\usepackage[displaymath,mathlines]{lineno}
\usepackage[authoryear,round]{natbib}
\usepackage[backref]{hyperref}
\usepackage{url}
\usepackage{upgreek}
\usepackage{comment}

\hypersetup{colorlinks=true,citecolor=cyan,urlcolor=cyan,linkcolor=blue}

\begin{document}


   \title{HelioCast: heliospheric forecasting based on white-light observations of the solar corona}
   \subtitle{I. Solar minimum conditions}
   
   \titlerunning{Heliospheric forecasting based on white-light observations}

   \authorrunning{Réville et al.}

   \author{V. Réville\inst{1}
          \and
          N. Poirier\inst{2,3,1}
          \and 
          A. Kouloumvakos\inst{4,1}
          \and
          A.P. Rouillard\inst{1}
          \and
          R.F. Pinto\inst{5}
          \and
          N. Fargette\inst{6,1}
          \and
          M. Indurain\inst{1}
          \and 
          R. Fournon\inst{1}
          \and
	      T. James\inst{1}
          \and
          R. Pobeda\inst{1}
          \and
          C. Scoul\inst{1}                     
          }

   \institute{IRAP, Université Toulouse III - Paul Sabatier, CNRS, CNES, Toulouse, France
         \and
         	Rosseland Centre for Solar Physics, University of Oslo, Postboks 1029 Blindern, N-0315 Oslo, Norway
	\and
		Institute of Theoretical Astrophysics, University of Oslo, Postboks 1029 Blindern, N-0315 Oslo, Norway
         \and 
      		The Johns Hopkins University Applied Physics Laboratory, 11101 Johns Hopkins Road, Laurel, MD 20723, USA
         \and
             	Département d'Astrophysique/AIM, CEA/IRFU, CNRS/INSU, Univ. Paris-Saclay \& Univ. de Paris, 91191 Gif-sur-Yvette, France
         \and
             	The Blackett Laboratory, Imperial College, London, United Kingdom}

  \abstract{We present a new 3D MHD heliospheric model for space-weather forecasting driven by boundary conditions defined from white-light observations of the solar corona. The model is based on the MHD code PLUTO, constrained by an empirical derivation of the solar wind background properties at 0.1au. This empirical method uses white-light observations to estimate the position of the heliospheric current sheet. The boundary conditions necessary to run HelioCast are then defined from pre-defined relations between the necessary MHD properties (speed, density and temperature) and the distance to the current sheet. We assess the accuracy of the model over six Carrington rotations during the first semester of 2018. Using point-by-point metrics and event based analysis, we evaluate the performances of our model varying the angular width of the slow solar wind layer surrounding the heliospheric current sheet. We also compare our empirical technique with two well tested models of the corona: Multi-VP and WindPredict-AW. We find that our method is well suited to reproduce high speed streams, and does --for well chosen parameters-- better than full MHD models. The model shows, nonetheless, limitations that could worsen for rising and maximum solar activity.}


   \maketitle

\section{Introduction}

Reliable space-weather predictions are essential to protect ground-based and spaceborne facilities, including manned missions to other planets. Yet, current models lack the accuracy necessary to make consistent and reliable predictions of all types of space weather events: flares, coronal mass ejections (CMEs) or high-speed streams (HSSs). This limitation is due to multiple factors. The physics of all relevant events is not fully understood. We cannot, for instance, fully anticipate which solar active region is going to flare, nor predict the eruption time and the properties of the resulting CME, which have strong consequences on our ability to predict the time of arrival of solar storms at Earth \citep{Riley2021}. The background propagation medium, i.e., the steady or ambient solar wind, also remains full of open questions. High-speed streams (HSSs) are produced in the low corona, and their exact formation mechanism are still debated. These velocity enhancements of the background solar wind further create co-rotating interaction regions (CIRs), where fast wind streams collide with slow wind streams forming structures that can be geoeffective \citep{GoslingPizzo1999, Yermolaev2012a}.

One way to circumvent the missing blocks of ab-initio models is to incorporate significant data assimilation in the models. A first very important input in most space weather models is related to the state of the Sun's magnetic field at a given time or during a time window. As a low beta plasma, the large-scale structure of the corona is indeed shaped by the magnetic field that we can measure through the Zeeman effect at the photosphere. The Wang-Sheeley-Arge model \citep{ArgePizzo2000, Arge2003} has developed and improved empirical relationships between the terminal solar wind speed and the properties of the 3D structure of the coronal magnetic field, obtained with a potential field source surface model \citep[PFSS,][]{Schatten1969, AltschulerNewkirk1969}. Nonetheless, precise measurements of the solar magnetic field required to drive PFSS models can only occur along the line-of-sight. Thus, usual synoptic “diachronic” magnetograms gather measurements taken at different times, and small bands at the central meridian are updated only once per Carrington rotation, while the rate of change of large-scale features relevant for space weather can be much faster. Moreover, at any given time, half of the solar surface cannot be constrained by any remote-sensing observation. Several techniques have been developed to address this problem. The National Solar Observatory Global Oscillation Network Group (NSO/GONG) uses helioseismology to gather information on the magnetic field on the far side of the Sun. The Air Force Data Assimilative Photospheric Flux Transport Model \citep[ADAPT,][]{Arge2010, Arge2013} uses GONG or HMI synoptic magnetogram along with flux transport models to assess what is the state of the solar magnetic field at a given time, yielding what we call “synchronic” magnetogram. 

However, significant differences exist between all available magnetograms. The most consistent source over a long period is obtained at the Mount Wilcox Solar Observatory (WSO), which exhibits scale dependent amplitude differences \citep{Virtanen2017} with other experiments, such as SOLIS and SDO/HMI. Amplitudes of the large-scale coronal magnetic field derived from the solar magnetograms are also difficult to reconcile with in situ measurements of the interplanetary magnetic field (IMF). This is known as the open flux problem \citep{Linker2017}, which states that the flux coming from regions of seemingly open magnetic fields (i.e. coronal holes) is not sufficient to account for the IMF open flux by roughly a factor 2. Although some possible solutions to the open flux problem have been proposed \citep{Riley2019b, Wang2022}, this illustrates how difficult it is to combine remote sensing and single point (or few points) measurements in the heliosphere to constraint solar wind models.

In this paper, we introduce an empirical technique to model the solar wind properties independently of any magnetogram source. We rely on remote-sensing observations of the solar corona in white-light obtained with the C2 coronagraph onboard the solar and heliospheric observatory (SOHO). We exploit tracking techniques of coronal regions of streamer maximum brightness \citep[SMB][]{Poirier2021}, i.e., the maximum brightness observed at a radius of $5 R_{\odot}$ in the plane perpendicular to the line of sight of SOHO. This method yields white-light (WL) maps updated at a rate of half a Carrington rotation and which contains information on the global state of the solar corona, including the far side of the Sun. We identify the SMB with the maximum of electron density, and with the position of the Heliospheric Current Sheet (HCS). From the position of the HCS, we directly access the structure of magnetic sectors. Moreover, following previous studies, we derive empirically the properties of the solar wind as a function of the angular distance of the solar wind plasma to the HCS \citep{Riley2001}. 

To test and assess the accuracy of this new empirical model of the solar corona, we propagate the obtained solar wind properties with a 3D MHD model from $0.1$ au to $1$ au. The combination of the WL boundary condition and the 3D MHD propagator forms the HelioCast model.  In section \ref{sec:mhd}, we present the 3D MHD model's equation and the characteristics of the simulations. In section \ref{sec:wl_bc}, we precisely describe how the ambient solar wind boundary condition is created from C2 WL measurements. We use OMNI and Ulysses data to relate the solar wind properties to the angular distance to the HCS at all latitudes. Section \ref{sec:comp1} compares the performance of the model depending on the variation of a single parameter $d$, the location of the transition from slow to fast wind going away from the HCS. Both point by point metrics and event based comparison are performed, allowing to assess the best value for the parameter $d$. Section \ref{sec:comp2} compares the accuracy of HelioCast with other types of boundary conditions, using the ab-initio models of the solar corona WindPredict-AW and Multi-VP. Finally, Section \ref{sec:limits} and Section \ref{sec:ccl} discuss the limits of the model and conclude our study.

\section{MHD heliospheric model}
\label{sec:mhd}

Throughout this work, we rely on the open source MHD code PLUTO \citep{Mignone2007} to perform numerical simulations of the inner heliosphere. The 3D MHD equations are solved in conservative form and can be written:

\begin{equation}
\label{MHD_1}
\frac{\partial}{\partial t} \rho + \nabla \cdot \rho \mathbf{v} = 0,
\end{equation}
\begin{equation}
\label{MHD_2}
\frac{\partial}{\partial t} \rho \mathbf{v} + \nabla \cdot (\rho \mathbf{vv}-\mathbf{BB}+\mathbf{I}p) = - \rho \nabla \Phi - 2 \rho \mathbf{\Omega_z} \times {\bf v} - \rho \mathbf{\Omega_z} \times \left({\bf e}_z\times {\bf r} \right),
\end{equation}
\begin{equation}
\label{MHD_3}
\frac{\partial}{\partial t} (E + \rho \Phi)  + \nabla \cdot [(E+p+\rho \Phi)\mathbf{v}-\mathbf{B}(\mathbf{v}\cdot \mathbf{B})] = 0,
\end{equation}
\begin{equation}
\label{MHD_4}
\frac{\partial}{\partial t} \mathbf{B} + \nabla \cdot (\mathbf{vB}-\mathbf{Bv})=0,
\end{equation}
where $\mathbf{v},\mathbf{B}$ are the velocity and magnetic field vectors, $p = p_{th}+ \mathbf{B}^2/2$ the total pressure, $\rho$ the mass density, $E$ the total energy and $\Phi$ the gravitational potential. The equations are solved in spherical coordinates $(r,\theta,\phi)$, in the rotating frame defined by the rotation vector $\mathbf{\Omega_z}$ normal to the ecliptic plane. We adopt a rotation frequency $\Omega_z=2.86 \times 10^{-6}$ s$^{-1}$, corresponding to the sidereal period of the Sun of $25.38$ days. The equations are closed by an ideal equation of state

\begin{equation}
    \rho e = \frac{p_{th}}{\gamma-1},
\end{equation}
where $e$ the internal energy. The use of a polytropic index $1 < \gamma < 5/3$, where $5/3$ is the value for a pure hydrogen gas, is a common technique to model fast electron thermal conduction in collisionless plasmas, at a very low computational expense \citep{Sakurai1985,KeppensGoedbloed1999,Matt2012,Reville2015a}. Indeed, heliospheric models do not require Spitzer-Härm collisional thermal conduction, making the system purely hyperbolic and much faster to converge. In this work, we will use a value of $\gamma=1.2$, for reasons explained in the following section. The code uses a constant radial resolution $\Delta r = 0.65 R_{\odot}$ and a fixed angular resolution of $1.875$ degrees. 

The main novelty of this model are the boundary conditions. The code has been modified to load and update time varying boundary conditions at any rate for all MHD variables. This makes physical sense because the inner boundary is located at $0.1 \mbox{ au} = 21.5 R_{\odot}$, where the solar wind is already (fast) super-alfvénic. Hence, all characteristics are pointing outward and all primitive variables can be set from the boundary conditions from which they will propagate in the computational domain. In the following section, we detail the method to build the solar wind solution at $0.1$ au.

\section{Ambient solar wind boundary conditions}
\label{sec:wl_bc}

\begin{figure*}
    \centering
    \includegraphics[width=6in]{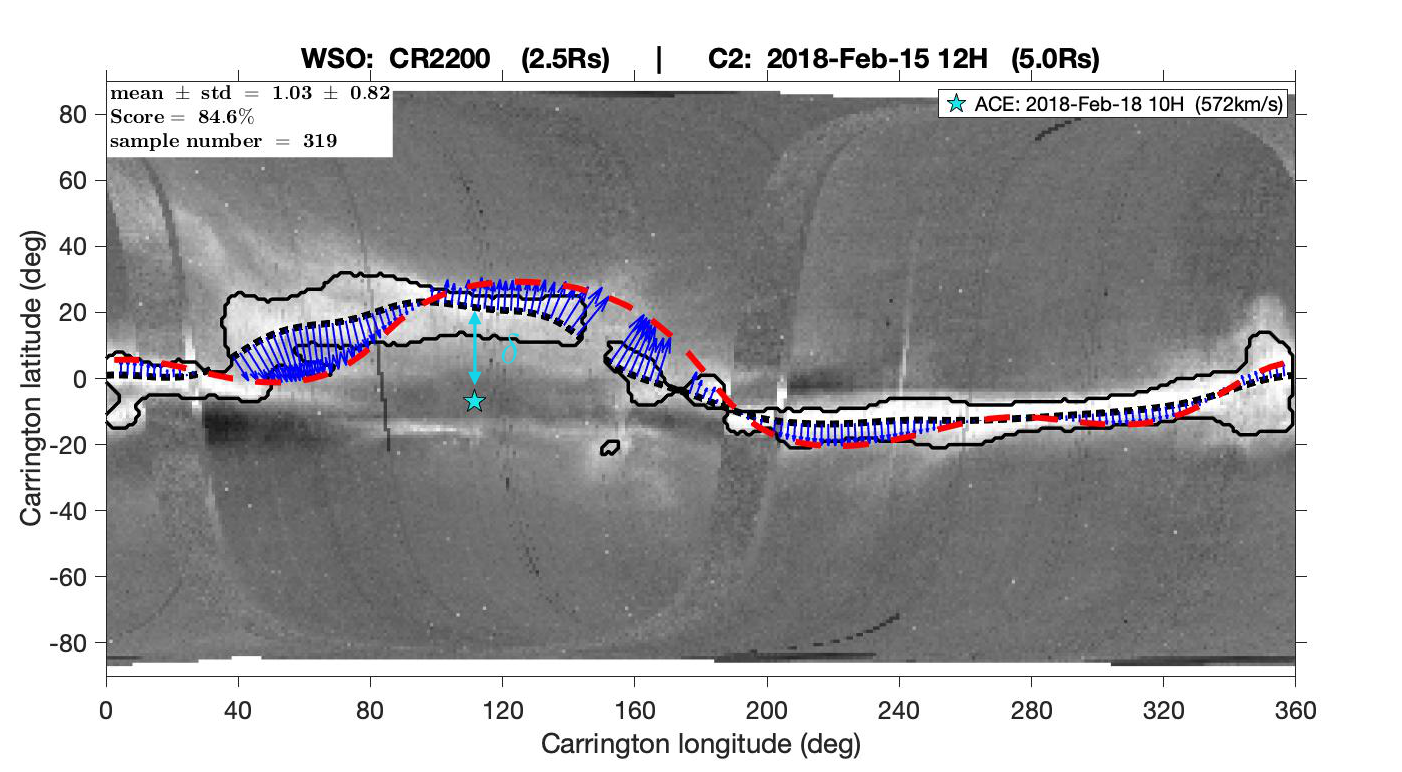}
    \caption{Example of the solar maximum brightness reconstruction for Feb. 15, 2018 from LASCO/C2 data. The SMB line is shown in dashed black, while the HCS predicted from the PFSS reconstruction is shown in dashed red. The agreement is good at this period with a score of $84.6$\%, computed with a distance-based metric that accounts for the streamer thickness \citep[see][for more details]{Poirier2021}. The blue star indicates the connectivity point of Earth (ACE), with the closest distance to the SMB $\delta$, displayed as a cyan arrow.}
    \label{fig:smb_wso}
\end{figure*}

White-light (WL) brightness measurements made by coronagraphs are direct indicators of the solar wind density. As such, they can be used to locate the heliospheric current sheet, which is known to harbor a slower and denser solar wind. In \citet{Poirier2021}, we used Carrington maps of the WL measurements made by LASCO/C2 \citep{Brueckner1995} to derive an optimization procedure of magnetic maps and potential field source surface parameters. In the present work, we wish to use WL data to deduce the solar wind properties in the corona, which will then be propagated in the heliosphere thanks to the MHD model, without any recourse to magnetic maps of the solar photosphere.

In Figure \ref{fig:smb_wso}, we show a particular example of the WL synoptic map on Feb, 15, 2018. Higher intensities are contoured by black curves, while the maxima are identified as the streamer maximum brightness (SMB) line, shown in dashed black. For comparison, we plot the HCS location predicted by a PFSS model based on the WSO magnetic map of CR2200. We see that both curves are close \citep[with a confidence score of 84.6\%, see][for more details]{Poirier2021}, and that we can use the SMB position as a proxy for the HCS. A statistical comparison with magnetic sector measurements made at 1 au further showed a good correlation between timings when the polarity switches sign and when the SMB is crossed \citep{Poirier2021}.

\begin{figure}
    \centering
    \includegraphics[width=4.5in]{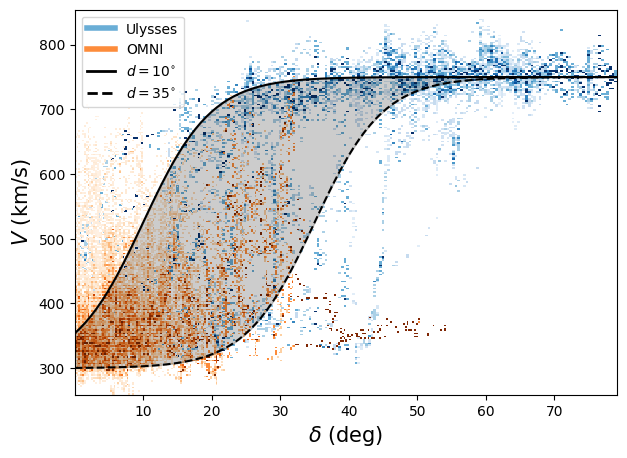}
    \caption{Relation between the solar wind speed (assumed constant beyond 1 au), and the angular distance to the HCS at 0.1 au $\delta$. The blue normalized histogram map comes from Ulysses latitudinal scan of 2007 while the orange comes from OMNI 1-hr averaged data for year 2018-2019-2020. The plain black and dashed black curves represent the logistic function described in equation \ref{eq:vdelta}, with values of $d=10$ and $d=35$, which seem to cover the most significant part of the transition between slow and fast wind streams.}
    \label{fig:vdelta}
\end{figure}

To build our solar wind model, we focus on the relation between the wind speed and the angular distance to the SMB. Figure \ref{fig:vdelta} shows the relation between the angular distance (in degrees) $\delta$ and the wind speed using two datasets. In blue, we show the computed 2D histogram of Ulysses data points, with 200 x 200 bins and normalized between 0 and 1 for each vertical $\delta$ bin, during the latitudinal scan of 2007. The distance $\delta$ to the HCS is computed using SMB reconstruction from LASCO/C2. In orange, we show the OMNI data histogram (taken at 1 au in 2018, 2019 and 2020) with the SMB computed with the same method. We see that the wind speed follows roughly a step function, going from $300$ km/s up to $750$ km/s. We assume that the solar wind has reached terminal velocity at 1 au and that it is not accelerated beyond (Ulysses orbit is between 1.3 and 2.5 au during this period).

\begin{figure}
    \centering
    \includegraphics[width=4.5in]{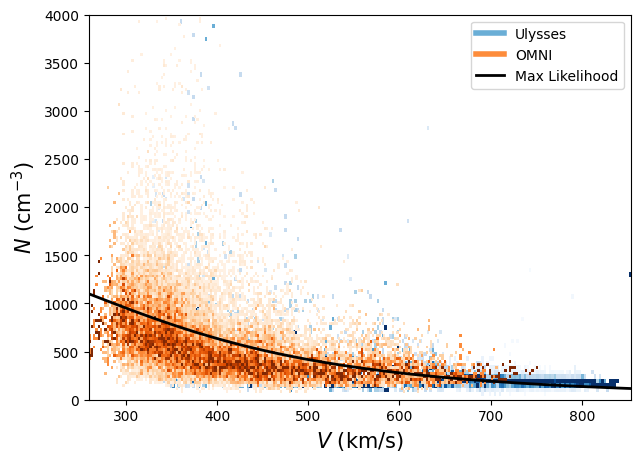}
    \caption{Proton density at 0.1 au as a function of the wind velocity, observed in OMNI and Ulysses data. The wind speed is assumed constant beyond 0.1 au. The density is assumed to decay as $r^{-2}$. }
    \label{fig:density}
\end{figure}

Figure \ref{fig:vdelta} shows also that there is a large variability or uncertainty in the position of the transition between slow and fast wind streams as a function of the angular distance $\delta$. In black, we plot two different models using the logistic function

\begin{equation}
    V(\delta) =  v_0+\frac{v_f-v_0}{1+e^{-r(\delta-d)}} \mbox{ km/s},
    \label{eq:vdelta}
\end{equation}

with 
\begin{equation}
    v_0=300 \mbox{ km/s},\; v_f = 750 \mbox{ km/s}, \; r=0.2 \mbox{ deg}^{-1}, \; d \in [10^{\circ}, 35^{\circ}],
\end{equation}
where $v_0$ is the solar wind speed at $\delta=0^{\circ}$, $v_f$ is the maximum solar wind, $r$ is the growth rate of the logistic function and $d$ is the location of the transition region. $v_0$ and $v_f$ are set based on the in-situ data of Ulysses and OMNI, while $r$ and $d$ can be fitted to the data. The two curves corresponding to $d=10^{\circ}$ and $d=35^{\circ}$ are represented in Figure \ref{fig:vdelta}. The black shaded region represents a range of  $d$ values between $10^{\circ}$ and $35^{\circ}$ for the $V(\delta)$ function. For the sake of simplicity, in the following the study, we fix $r=0.2$ deg$^{-1}$ and assess the performance of the model according to different values of $d \in [10^{\circ}, 35^{\circ}]$. For each model / $d$ value, we want to express the remaining hydrodynamical quantities as a function of the wind speed at 0.1 au only. We thus look at the same data set of OMNI and Ulysses measurements, assuming a $r^{-2}$ decay for the density starting at $0.1$ au. The result is shown in Figure \ref{fig:density}, where the density shows a clear decreasing trend with wind velocity, which is consistent with the well known fact that the mass flux is approximately constant in the solar wind \citep[see][]{Wang2010}.

We thus chose a classical decreasing function to model the density dependence on $V$, with a relation of the form:
\begin{equation}
    N(V) = \frac{N_0}{1 + \left(\frac{V}{V_0}\right) ^\alpha}
    \label{eq:ndelta}
\end{equation}
where $N (V)$ is the modeled density and $N_0$, $V_0$ and $\alpha$ are parameters to fit to the data. We assume a white noise model with a $\sigma$ = 30 cm$^{-3}$ dispersion, as well as independent measurements. The likelihood function we seek to maximize is then given by: 
\begin{equation}
    \mathrm{p}(N~|~V, \mathbf{\uptheta}) = \displaystyle \prod_{i} 
    \frac{1}{\sqrt{2 \pi \sigma^2}} e^{-\displaystyle~\frac{1}{2 \sigma^2} |~N_i(V_i) -N(V_i, \mathbf{\uptheta})~| ^2}\\
\end{equation} with $\mathbf{\uptheta} = [N_0, V_0, \alpha]^T$ the parameter vector to fit. 
The maximum likelihood is obtained for $N_0$=1600~cm$^{-3}$, $V_0$ =345~km/s and $\alpha$ = 2.8.

\begin{figure}
    \centering
    \includegraphics[width=6in]{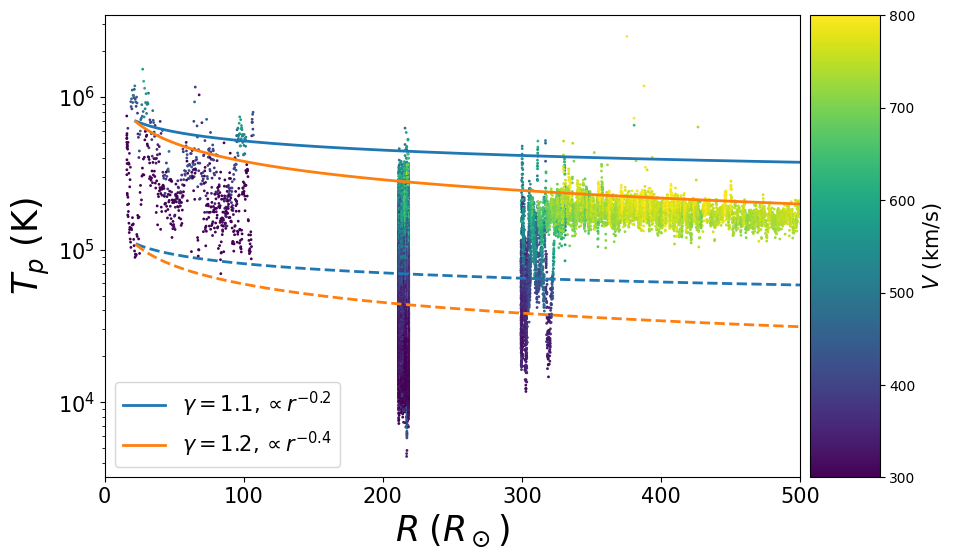}
    \caption{Proton temperature as a function of the distance to Sun. The three groups of points represent PSP E8 data, OMNI data (2018-2019-2020) and Ulysses data (2007). Points are colored as a function of the wind speed, with slow wind (300 km/s) in dark purple and fast wind (800 km/s) in yellow. Polytropic decay profiles are shown for the fast and slow winds and for $\gamma=1.1, 1.2$, and base temperatures of $7 \times 10^5$ K (plain) or $1.1 \times 10^5$ K (dashed) at 0.1 au.}
    \label{fig:temp_rad}
\end{figure}

For the wind proton temperature, more discussion is necessary, as its evolution with distance to the Sun is complex and still not completely understood. In Figure \ref{fig:temp_rad}, we present the proton temperature of the two datasets previously used, as well as the proton temperature measured by Parker Solar Probe (PSP) during encounter 8, between $16$ and $100$ solar radii. The PSP proton temperature correspond to the L3 product of the SWEAP/SPAN-i obtained through the moment of the velocity distribution function \citep[see][]{Kasper2016, Verniero2022}. The colorscale of the scattered points represent the wind speed from 300 km/s (dark purple) to 800 km/s (yellow). The temperature profile in the solar wind has been extensively discussed in the literature, with different decay properties depending on the species, range of radial distance and parallel and perpendicular direction with respect to the magnetic field. \citet{Hellinger2011} used Helios proton data and found that between 0.3 and 1.0 au the global temperature was proportional to $r^{-0.74}$. For the electrons, Helios data suggests that a decay exponent between $-0.3$ and $-0.7$ is likely compatible with most of the observations, depending on the wind velocity \citep{Stverak2015}.

We consider here a polytropic MHD model which imposes a certain decay of the temperature with distance, depending on the value of $\gamma$. Following the polytropic relation, $T \propto \rho^{\gamma-1}$, and thus

\begin{equation}
    T \propto r^{-2(\gamma-1)},
\end{equation}

assuming a $1/r^2$ decay of the wind density. It is unlikely that a single power law and thus a single polytropic index value is able to accurately reproduce the temperature profile of the solar wind. In Figure \ref{fig:temp_rad}, we plot two typical temperature decay profiles for $\gamma=1.1$ and $\gamma=1.2$ for two different temperatures at the inner boundary condition. These values roughly bound the temperature distribution at 0.1 au. For the slow wind, both values of $\gamma$ could be consistent with the data. However, for the fast wind, we see that only the $\gamma = 1.2$ curve matches the asymptotic values of the temperature around 2 au. Thus, we will use this value within the MHD model, and we assume a typical $\propto r^{-0.4}$ decay of the solar wind temperature, with distance to the Sun.
\begin{figure}
    \centering
    \includegraphics[width=4.5in]{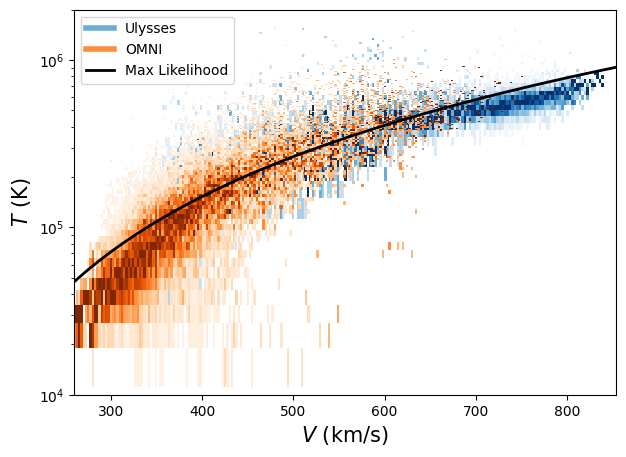}
    \caption{Proton temperature at 0.1 au as a function of the wind velocity, observed in OMNI and Ulysses data. The wind speed is assumed constant beyond 0.1 au. The temperature is assumed to decay as $r^{-0.4}$.}
    \label{fig:temperature}
\end{figure}

In Figure \ref{fig:temperature}, we show the temperature, brought back to 0.1 au, as a function of the wind speed. The data shows an increasing trend that we fit with the following function:
\begin{equation}
    T(V) = (a*V-b)^2
    \label{eq:tdelta}
\end{equation}
where $a$ and $b$ are parameters to fit to the data and $V$ is expressed in km/s. We use a similar approach to the one presented for the density, and find a maximum likelihood for $a$ = 1.235~K$^{0.5}$~s~km$^{-1}$ and $b$ = 103 K$^{0.5}$.

The temperatures chosen at 0.1 au for the orange curves in Figure \ref{fig:temp_rad}, thus correspond, according to this law, to velocities of 350 km/s ($1.1 \times 10^5$ K) and 750 km/s ($7 \times 10^5$ K) for the slow and fast wind respectively. 

\begin{figure}
    \centering
    \includegraphics[width=6.5in]{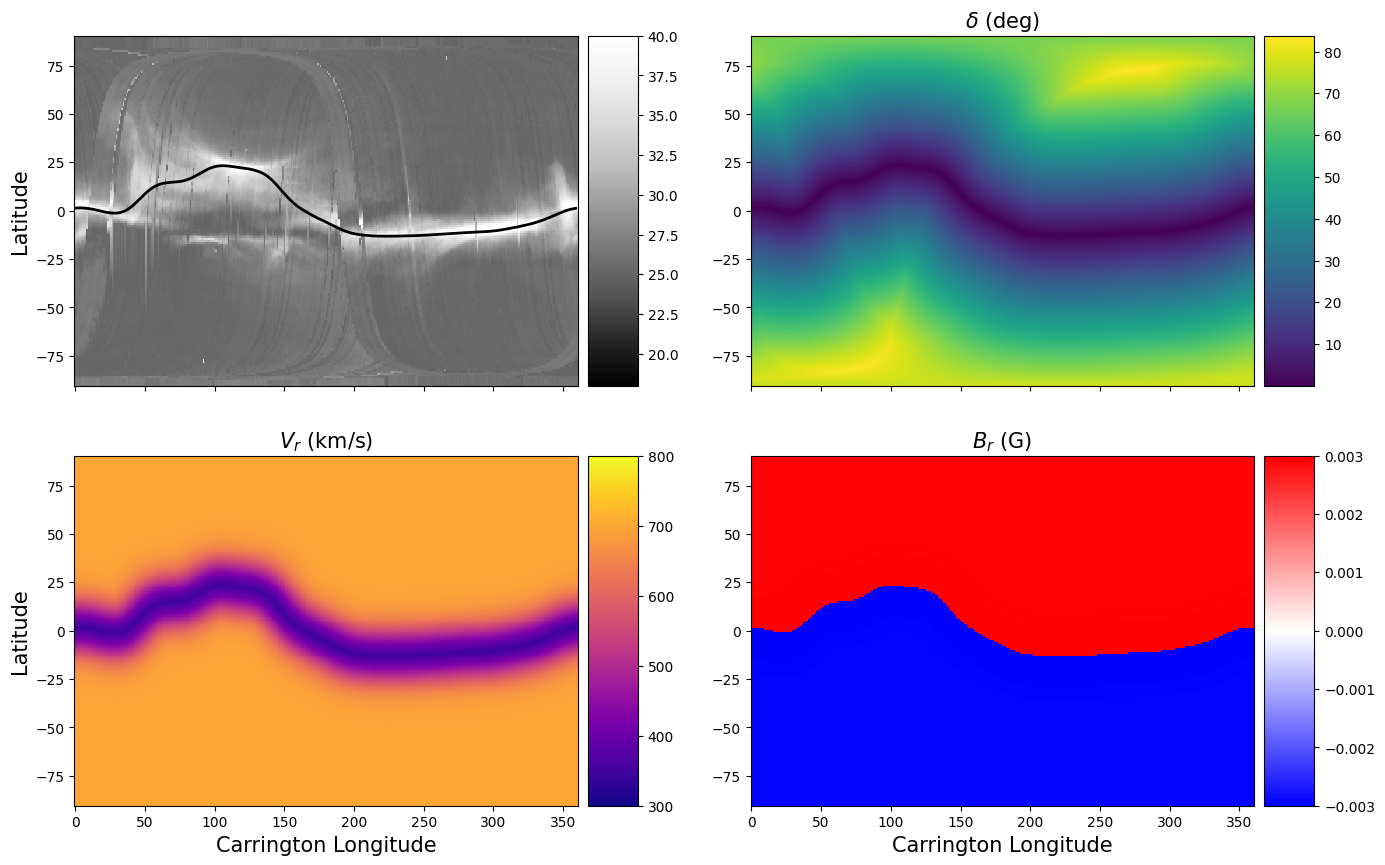}
    \caption{Example of a boundary condition derived from the SMB of February 15, 2018. The HCS is identified with the SMB (top left panel), from which we derive an angular distance $\delta$ (top right), and a wind velocity (bottom left, with $d=10$). The magnetic flux is evenly distributed on each side of the HCS/SMB, in the $B_r$ component (bottom right).}
    \label{fig:smb_to_wind}
\end{figure}

Finally, the last input is the magnetic field that we assume purely radial at 0.1 au. As the main purpose of this study is to avoid the use of observed solar magnetogram, we rely on the fact that the magnetic flux is homogenized in the heliosphere, as shown with Ulysses \citep{Smith2011}. \citet{Reville2017} have shown with MHD simulation, that the homogenization process through latitudinal Lorentz forces is accomplished by $10R_{\odot}$, and thus at the inner boundary (at $21.5 R_{\odot}$), we report the typical flux observed at 1 au, $\Phi / (4 \pi) = 3nT $au$^{2}$ \citep[see, e.g.][]{Badman2021}, homogenized on each hemisphere, separated by the Heliospheric Current Sheet.

In Figure \ref{fig:smb_to_wind}, we illustrate the production process of the inner boundary condition, starting with the white-light map of 2018 Feb. 15, shown in Figure \ref{fig:smb_wso}. The top panel represents the white-light map with the computed SMB line (note that the numerical values are on an arbitrary scale). The remaining panels show $\delta$, $V_r$ and $B_r$ computed from the SMB. Density and temperature are computed as equations \ref{eq:ndelta} and \ref{eq:tdelta}. The transverse components of the velocity and magnetic field are set to zero.

\section{Validation on the first semester of 2018}
\label{sec:comp1}

\subsection{Point by point comparison}

In the following section, we compare the results of our model, HelioCast, with data of the solar wind measurements made at 1 au. We focus on the first semester of 2018, as this is a well studied interval \citep[see, e.g.][]{Samara2021}, with many HSSs observed in the ecliptic plane, lasting for several days. This period is close to solar minimum, and only a few interplanetary mass ejections have been observed at Earth\footnote{See, https://izw1.caltech.edu/ACE/ASC/DATA/level3/icmetable2.htm}. All were travelling at relatively slow speed ($\sim 400$ km/s), on timescales $\leq 1$ day. Hence, we did not perform any CME removal on the data.

\begin{figure}
    \centering
    \includegraphics[width=6in]{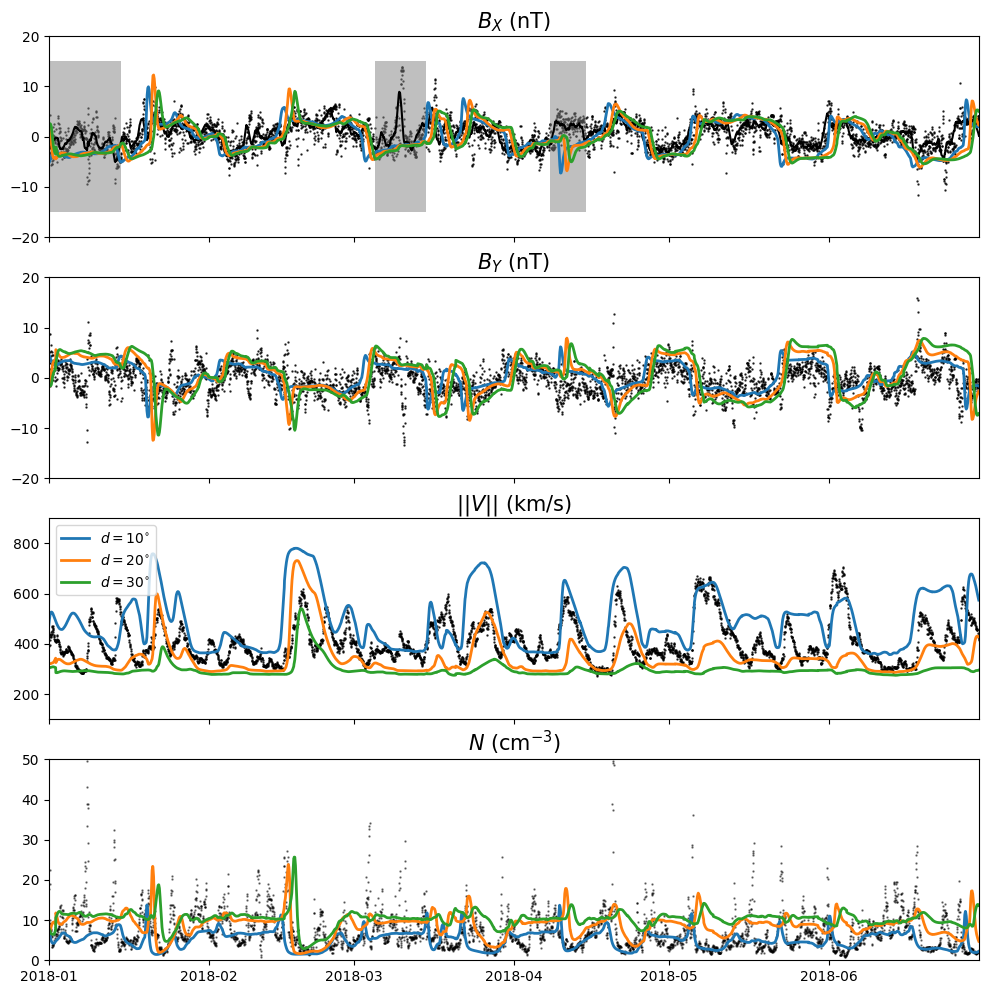}
    \caption{Plasma parameters at Earth (OMNI), compared with the results of HelioCast with three different value for the parameter $d$. Magnetic field components X and Y (GSE coordinates) are well reproduced. The black line in the top panel is a one-day running average of the $B_X$ data. HSSs are reproduced with varying amplitudes depending on $d$ in the third, velocity panel. The bottom panel shows the density variations, which are best reproduced for $d=10$.}
    \label{fig:in_situ_hc}
\end{figure}

Figure \ref{fig:in_situ_hc} shows the comparison between three HelioCast runs with $d=10, 20, 30$. We display two components of the magnetic field $B_X, B_Y$ in the GSE coordinates system. The X component is opposite to the radial component in the spherical coordinates system and corresponds to the polarity of the interplanetary magnetic field. The X component of the magnetic field is well reproduced by the model, for all $d$ values, which play no role in the magnetic field initialization. Small differences between the curves can be understood as the evolution, through the MHD model, of pressure equilibria around the current sheet.

In the second panel of Figure \ref{fig:in_situ_hc}, we show the Y component of the magnetic field. The azimuthal magnetic field is set to zero at 0.1 au, and thus this component is mostly a consequence of the IMF Parker spiral. Interaction between slow and high-speed streams could also play a role in the dynamical evolution of $B_X$ and $B_Y$. We observe again a good agreement between the different realization of the model and the data, with more differences between different $d$ values. Notably, the $B_Z$ component is very weak in all the models, and as such HelioCast is not, at the moment, a useful tool to predict this parameter.

In the third panel of Figure \ref{fig:in_situ_hc}, the data display more than 20 HSSs that we aim at predicting with HelioCast. In contrast with the magnetic field, we do see large differences in the model solutions depending on the width $d$ of the transition from slow to fast solar wind away from the HCS/SMB. The amplitude of the maximum velocity, the value of the low speeds and to some extent the occurrence of the peaks is affected by the choice of $d$. For $d=10^{\circ}$, most HSSs are detected, but their amplitude is too high at the beginning of the interval, while improving over the six-month period. The slow wind speed predicted by the model is also too high, around 400 km/s, when the data shows a slow wind plateau around 300 km/s. Conversely, for $d=30^{\circ}$, the Earth stays most of the time in  300 km/s slow wind, and only a few high-speed streams do emerge in the solution. Finally, the intermediate value $d=20^{\circ}$, shows a better agreement for the slow / fast wind amplitudes as well as a good number of HSS predictions. As shown in the last panel of Figure \ref{fig:in_situ_hc}, the wind density performance is strongly correlated with the velocity and an over (under) estimation of the density correspond to an under (over) estimation of the wind speed. For $d=10^{\circ}$, the density variation seems reasonably reproduced.

For the rest of the study, we will focus on the prediction of the wind velocities and the occurrence of high-speed streams. To go further in our analysis, we must define quantitative metrics to assess the performances of our models. In the past few years there have been a number of works and discussion on the right way to assess the validity of a model, in terms of forecasting performance \citep[see][]{Owens2018}. Comparing two time series, the most basic idea is to perform usual analysis such as computing the standard deviation (SD), the root-mean-square error (RMSE) or the correlation between the time series \citep[see][for more advanced methods]{Samara2022}. As we are interested in forecasting HSSs, we compare the wind velocity time series and give the value of the corresponding metrics in Table \ref{tab:taylor}. 

\begin{table}[]
    \centering
    \begin{tabular}{l | c |  c | c  |}
        Model & SD (km/s) & RMSE (km/s) & R \\
        \hline
        $d=10^{\circ}$ & 112 & 130 & 0.52\\
        $d=15^{\circ}$ & 99 & 96  & 0.45\\
        $d=20^{\circ}$ & 76 & 107 & 0.36\\
        $d=25^{\circ}$ & 52 & 124 & 0.29\\
        $d=30^{\circ}$ & 32 & 136 & 0.25\\
        \hline
        Multi-VP & 59 & 106 & 0.04\\
        WP-AW & 72 & 99 & 0.3\\
    \end{tabular}
    \caption{Point-by-point metrics comparison between models for various $d$ values, as well as Multi-VP and WindPredict-AW (WP-AW) boundary setup. RMS values are all close to $100$ km/s, without any clear hierarchy. Pearson correlation coefficient shows that best models are obtained for $d=10/15^{\circ}$.}
    \label{tab:taylor}
\end{table}

The root-mean-square error (RMSE) is defined by:

\begin{equation}
    \mbox{RMSE} = \sqrt{ \frac{1}{n} \sum^{n}_{k=1} (f_k-o_k)^2},
\end{equation}
where $f_k$ and $o_k$ represent discretized instances of the forecasted and observed samples, respectively, and $n$ is the number of samples taken along the time series. The standard deviation (SD) is computed using the RMSE with respect to the average forecasted (or observed) constant signal. It represents the typical variation of the signal and is crucial information to analyze the RMSE. For instance, in Table \ref{tab:taylor}, we see that the RMSE yield similar values for most values of $d$. Yet, the standard deviation clearly decreases, because the amplitude of the wind speed variation are much less for large values of $d$. Hence, for values of $d>25^{\circ}$, the discrepancies between the SD and RMSE values seem to indicate that models are doing poorly, which is what we naturally see by eye in Figure \ref{fig:in_situ_hc}.

Another way of comparing times series is to compute the Pearson correlation coefficient (R), defined as follows:

\begin{equation}
    R = \frac{\sum^n (f_k - m_f) (o_k -m_o)}{\sqrt{\sum^n (f_k - m_f)^2 (o_k -m_o)^2}},
\end{equation}
where $m_f$ and $m_o$ are the average value of the forecasted and observed signal, respectively.

\begin{figure}
    \centering
    \includegraphics[width=4.5in]{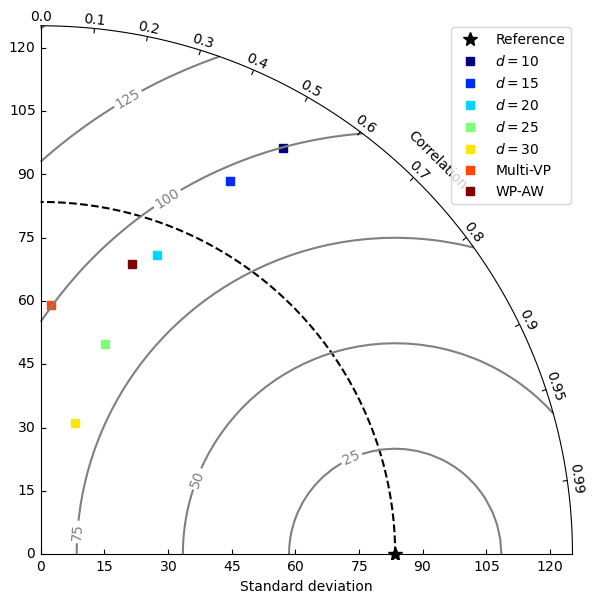}
    \caption{Taylor diagram comparing solar wind speed time series for the various models. The dashed line correspond to the data's standard deviation of 83 km/s. Models are then plotted as a function of their SD and their correlation coefficient (see Table \ref{tab:taylor}). Best models are the closest to the star along the dashed line.}
    \label{fig:taylor}
\end{figure}

According to the numbers in Table \ref{tab:taylor}, the correlation coefficients are a pure decreasing function of $d$, which would mean that the model with $d=10^{\circ}$ is the one that performs best. Note, that the p-value associated with the Pearson coefficient are all extremely low $<10^{-10}$, which means that the probability of obtaining these kinds of correlation by chance is very weak. \citet{Taylor2001} has proposed a way to gather all these indicators in a single diagram. In Figure \ref{fig:taylor}, we report such a diagram where the different models are classified according to their SD and Pearson correlation coefficient R. As shown by \citet{Taylor2001}, there exists a geometrical relation between the RMSE, the SD and R, which stands that if the star is reporting the value of the reference standard deviation (i.e. of the data sample), the distance to the star is related to the RMSE. Hence, the best models are the ones with the highest correlation coefficient and the closest to the dashed black line. In our case, we find that the model with $d=20^{\circ}$ is doing much better than $d=10^{\circ}$ and $d=30^{\circ}$, which is expected from Figure \ref{fig:in_situ_hc}. 

\begin{table*}[]
    \centering
    \begin{tabular}{l | c |  c | c  | c | c | c | c | c|}
        Model & TP & FP & FN & POD & TSS & Hiedke SS & \% overlap  av. & av. delay (hours) \\  
        \hline
        $d=10^{\circ}$ & 13 & 2 & 13 & 0.5  & 0.36 & 0.32 & 92 & -56 \\
        $d=15^{\circ}$ & 16 & 2 & 10 & 0.61 & 0.49 & 0.45 & 69.5 & 7.75 \\
        $d=20^{\circ}$ &  7 & 0 & 19 & 0.27 & 0.27 & 0.13 & 62 & 23.5\\
        $d=25^{\circ}$ &  3 & 0 & 23 & 0.11 &  0.115 & 0.03 & 65 & 24.5 \\ 
        $d=30^{\circ}$ &  1 & 0 & 25 & 0.04 &  0.04 & 0.006 & 74 & 21.0 \\
        \hline
        Multi-VP& 13 & 2 & 13 & 0.5 & 0.36 & 0.30 & 91 & -69 \\
        WP-AW &  11 & 5 & 15 & 0.42 & 0.19 & 0.19 & 83 & -9.2 \\
    \end{tabular}
    \caption{Event based comparison between models for various $d$ values, as well as Multi-VP and WindPredict-AW boundary setup.}
    \label{tab:events}
\end{table*}

\subsection{Event based comparison}

Space weather models can also be evaluated with other types of metrics. \citet{Owens2018} discussed the various advantages and disadvantages of the method described before with regard to event-based methods. Event based comparisons focus on the ability of a given model to predict a number of events characterized by some properties. For instance, as we are interested in HSSs, we can define an HSS event as a wind velocity increase above some threshold. This technique attempts to match predicted and observed events and then assess the performance of the models. 

We define an HSS event in the OMNI data as a continuous period of more than 12h, where the wind speed is above 400 km/s from January 1st to June 30th of 2018. Following \citet{Reiss2016}, we then construct a list of events for the observed data and each of the HelioCast computed models. We then match, when possible, every event in the model with a single event in the data. We first chose the centered time interval in the observed HSS events, and then ask that there is at least some overlap between the modelled and observed event. If two modelled events refer to the same observed one, we merge the two and the resulting overlap. We can then compute the number of true positive (TP, when an HSS is predicted correctly by the model), false positive (FP, when an HSS is predicted but not observed) and false negative (FN, when an observed HSS is not predicted) \citep[see, e.g.,][for more details]{Reiss2016}. 

In Table \ref{tab:events}, we gather the results for all HelioCast models (as well as the MHD models introduced in the next section). Many measures can be extracted from the value of TP, FP, and FN. As examples, we compute the probability of detection 

\begin{equation}
    \mbox{POD} = \frac{\mbox{TP}}{\mbox{TP + FN}},
\end{equation}
the true skill score 

\begin{equation}
    \mbox{TSS} = \frac{\mbox{TP}}{\mbox{TP + FN}} - \frac{\mbox{FP}}{\mbox{FP + TN}},
\end{equation}
where TN is the number of true negatives, i.e., the number of times that non detection have been detected. The TSS compares the difference between the probability of detection and the probability of non detection. Finally, we also compute the Hiedke skill score
\begin{equation}
    \mbox{Hiedke SS} = \frac{\mbox{2 ((TP $\times$ TN) - (FN $\times$ FP))}}{\mbox{P ( FN + TN) + N (TP + FP)}},
\end{equation}

where  P = TP + FN and N = FP + TN. A Hiekde skill score of 0 means that the forecast does no better than a random one, and can go as high as unity. All these scores are listed in Table \ref{tab:events}. They show a clear hierarchy of the models and it is clear that it is the model using $d=15^{\circ}$ that performs best. This differs slightly from the previous analysis (results reported in Figure \ref{fig:taylor}), where the model for $d=20^{\circ}$ was found to better perform.

\begin{figure*}
    \centering
    \includegraphics[width=6in]{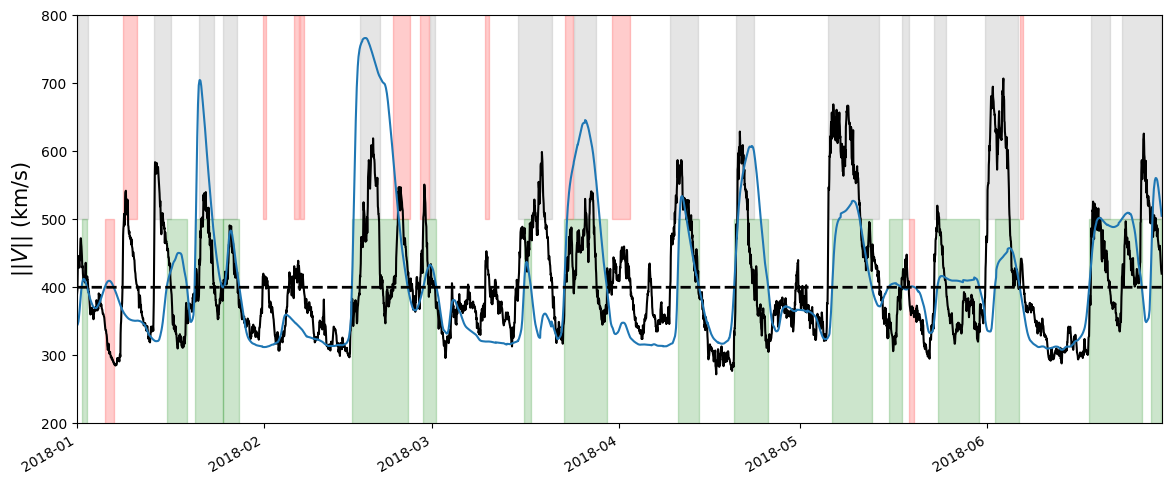}
    \caption{Event based comparison between HelioCast $d=15^{\circ}$ model (blue) and OMNI data (black) for the total wind speed. Shaded regions in the bottom part of the plot represent detected events in the model, true positives when green / false positives when red. The threshold for the HSS detection is marked by the black dashed line. The top shaded regions are the HSSs detected in the data, and red color indicates the false negatives.}
    \label{fig:event_compar}
\end{figure*}

As additional measures, we compute the average overlap percentage in the model and the average delay of all true positives in the model.  In Figure \ref{fig:event_compar}, we illustrate the event selection and matching procedure for $d=15^{\circ}$.  We see that most events predicted by the model are real HSSs, and that there is not much delay in general for the arrival of the HSS at Earth. The amplitude of the speed enhancement is also reasonably reproduced, although this is not measured in any metric of the event based comparison. 

There is, of course, some fine structure in the observed HSS that is not reproduced in the model. Also, some non-detections are influenced by the value of the velocity threshold set for the algorithm. For example, one of the false negatives in early March, could have been predicted with a lower value of the threshold. Our WL based boundary condition does nonetheless provide good results, and it shall be interesting to know how it performs compared to other, much more complex models of the inner heliosphere.

\section{Comparison with other models}
\label{sec:comp2}

Using the SMB as a single proxy for all properties of the solar wind is a strong simplification, and it is thus of primary importance to compare this model with more realistic solar wind models. We will be using two additional models, WindPredict-AW and Multi-VP, which yield all the global MHD quantities, namely the wind speed, density and the interplanetary magnetic field from the solar surface up to a few tens of solar radii. 

WindPredict-AW is a full global MHD model of the inner heliosphere, which includes Alfvén wave turbulent phenomenology as the main driver of coronal heating. It has been successfully validated against in situ data, between $0.17$ and $0.5$ au, comparing with the measurements of the first Parker Solar Probe perihelion \citep{Reville2020ApJS}. \citet{Parenti2022} have validated the model against remote sensing white-light and extreme UV measurements (SOHO/LASCO, K-Cor, SDO/AIA). Moreover, the model has been shown to reproduce the dynamics of flux ropes created at the tip of helmet streamers \citep{Reville2020ApJL, Reville2022}.

WindPredict-AW uses as main input solar magnetograms, which determine the distribution of the solar magnetic field at any given time. We use a spherical harmonics decomposition of the observed radial field $B_r$, and reconstruct a potential field solution up to the harmonic degree $l=15$ for the initialization of the simulation. The structure of the magnetic field is then maintained in the boundary conditions \citep[see,][for more details on the boundary conditions]{Parenti2022}. The second important parameter of the model is the transverse motion amplitude $\delta v$, which is chosen to be constant all over the solar surface and equal to $12$ km/s. This value will determine the input Alfvén-wave Poynting flux $\rho v_A \delta v^2$ at the surface. The value chosen here is similar to previous works \citep{Reville2020ApJS, Reville2022}. The computational domain extends up to $130 R_{\odot}$ with an angular resolution of 2 degrees. 

Multi-VP is a multiple flux tube model \citep{Pinto2017}, which uses a PFSS extrapolation to get the coronal magnetic structure and then runs 1D hydrodynamical simulations along all flux tubes determined by a given angular resolution. We use a 5x5 degrees of angular resolution on the full  $4\pi$ sphere, which gives a total of 2592 flux tube computations. Given the reduced resolution and the multi-flux tube character of Multi-VP, the runs are computationally much cheaper and are thus a good middle ground between the full MHD runs of WindPredict-AW and the empirical WL solar wind solution presented in this paper. Note that Multi-VP can be run at higher resolution \citep[e.g., 2x2 deg, see][]{Poirier2020ApJS}. 

We performed six simulations with each model covering the first semester of 2018. We thus chose one magnetogram per Carrington rotation, going from CR number 2200 to CR 2205. We used ADAPT magnetograms of the photospheric magnetic flux \citep{Arge2010, Arge2013}. They come in 12 different realizations, depending on the properties of the flux transport model, and we chose to use the first realization for each magnetogram. We use the solution obtained in both models at $0.1$ au and use it as a boundary condition for our heliospheric MHD solver, to  ensure consistency in the comparison. 

\begin{figure*}
    \centering
    \includegraphics[width=6in]{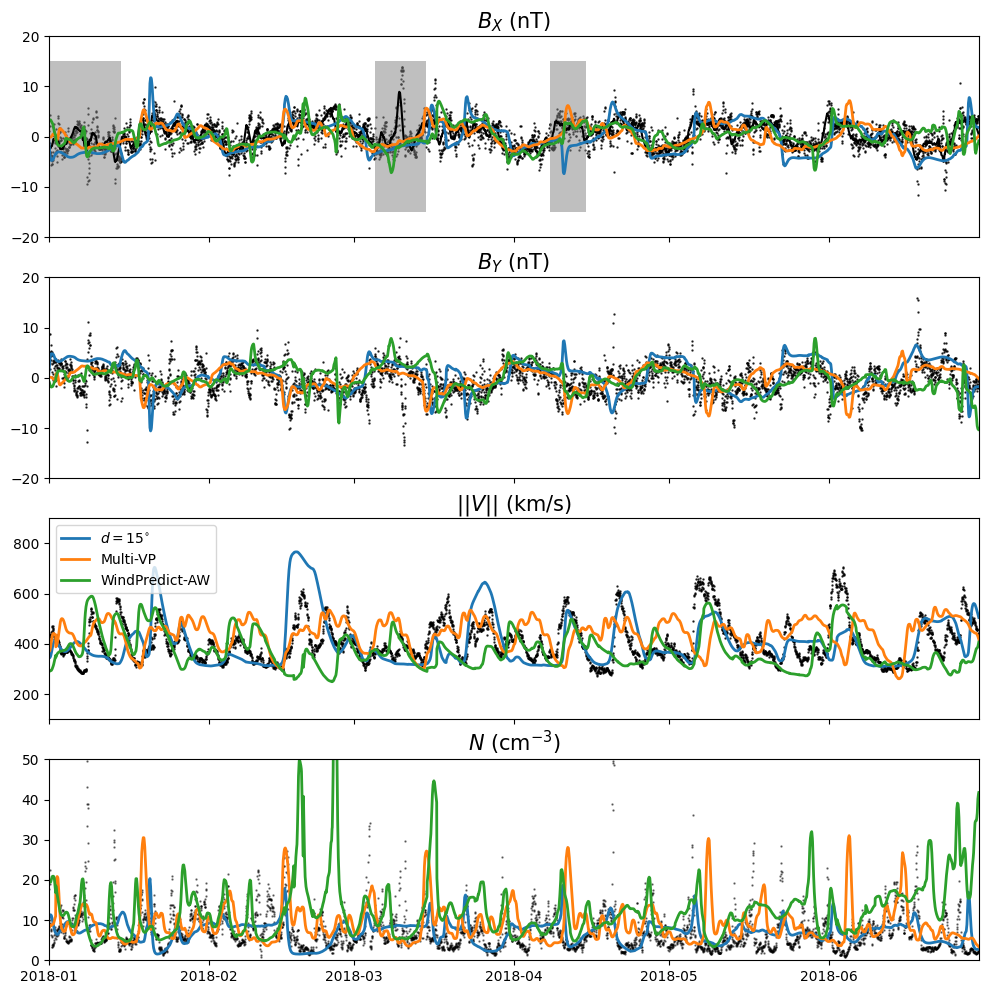}
    \caption{Plasma parameters at Earth (OMNI, same as Figure \ref{fig:in_situ_hc}), compared with the results of HelioCast ($d=15$), Multi-VP and WindPredict-AW. Grey regions indicate where the polarity of the magnetic field is not correctly captured by HelioCast (see section \ref{sec:limits}).}
    \label{fig:in_situ_mvp_wp}
\end{figure*}

In Figure \ref{fig:in_situ_mvp_wp}, we show the comparison of the in-situ data, the best model obtained with HelioCast, and the results obtained with Multi-VP and WindPredict-AW.  We highlighted three gray zones where the polarity is not predicted by the HelioCast model. This is because the SMB does not correspond to the HCS in those places. Pseudo-streamers are wrongly identified as helmet streamers and lead to the wrong polarity initialization in the model. We discuss this limitation extensively in section \ref{sec:limits}. The results of the point-by-point metrics and event analysis have been added to Table \ref{tab:taylor}, Table \ref{tab:events} and Figure \ref{fig:taylor}. 
It is here interesting to compare the different diagnostics. In the Taylor diagram of Figure \ref{fig:taylor}, the WindPredict-AW simulations are close to the HelioCast model $d=20^{\circ}$, which is the best performing model according to this metric. The shape of the velocity signal is indeed similar to the results obtained with HelioCast and the amplitude of HSSs remains in good agreement with the data (in contrast with HelioCast). Multi-VP on the other hand, conveys more structure to the velocity profiles, with smaller variations on each HSS. This is likely due to the multi-flux tube character of the simulation that does not account for interaction between adjacent flux tubes. There is nonetheless similar fine structuring in the data at some locations. 
In terms of event-based analysis, the simulations based on Multi-VP are doing slightly better than with WindPredict-AW, except for the average delay. One can see that in general, Multi-VP predicts very consistently the amplitude of HSSs, but does predict longer events than the other models (and the data). With both models, the probability of detection still remains below HelioCast ($d=15^{\circ}$). 

\section{Limits of the model, pseudo-streamers and path to solar maximum}
\label{sec:limits}

The identification process of the HCS/SMB from WL observations raises several issues, which have already been discussed in detail in \citet{Poirier2021}. One of them has been particularly noticed in Figure \ref{fig:in_situ_mvp_wp}, where a wrong estimate of the location of the HCS/SMB leads to an incorrect prediction of the magnetic sector at 1 au. 

\begin{figure}
    \centering
    \includegraphics[width=4.5in]{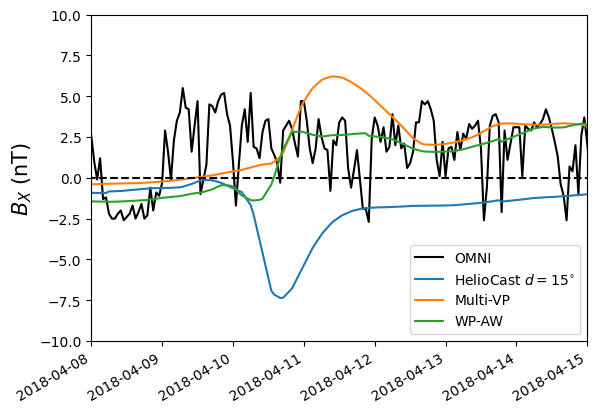}
    \caption{Zoom on the $B_X$ component of Figure \ref{fig:in_situ_mvp_wp} for the third gray region. We see that despite some delay, the two magnetogram based models correctly capture the IMF polarity while HelioCast does not.}
    \label{fig:bx_april}
\end{figure}

\begin{figure}
    \centering
    \includegraphics[width=6in]{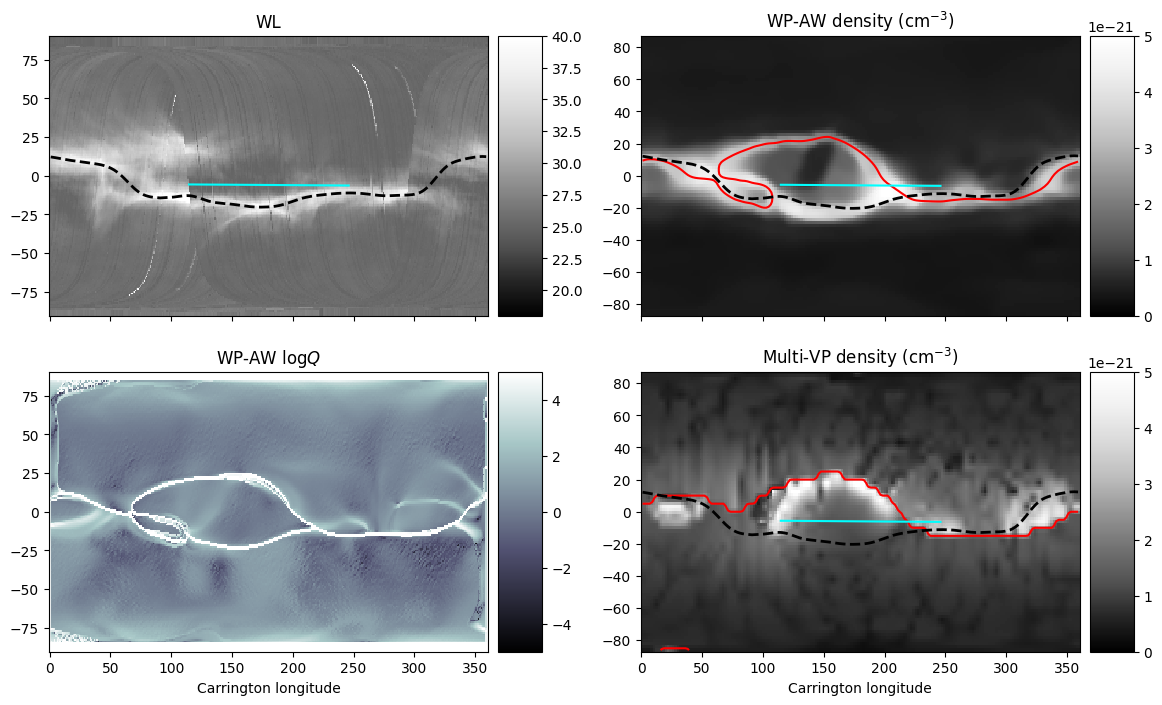}
    \caption{Comparison of the density structures, position of the HCS and squashing factor. The top left panel shows the WL map obtained on March 20th and the selected SMB with our technique. The two right panels show the density at 0.1 au obtained with Multi-VP and WP-AW, with the actual HCS in red for CR 2202. The projected trajectory of Earth is plotted in blue. The bottom left panel is the logarithm of the squashing factor of the WP-AW simulation.}
    \label{fig:qsls}
\end{figure}

Figures \ref{fig:bx_april} and \ref{fig:qsls} illustrate this process. Figure \ref{fig:bx_april} is a zoom on the third gray period highlighted in Figure \ref{fig:in_situ_mvp_wp} between March, 4 and March 15, 2018. We see that the polarity of the IMF predicted by HelioCast is mostly wrong for the whole interval, while it is captured (with some delay) in the Multi-VP and WindPredict-AW based simulations. The reason for this incorrect behavior is shown in Figure \ref{fig:qsls}. The results of the WL selection of the SMB is shown in the top left panel. Then, we plot the density obtained in Multi-VP and WindPredict-AW for the corresponding period, along with the true HCS in red. We clearly see that, in the middle region, the SMB differs from the HCS. Our algorithm picked the bottom part of the density arc while the HCS actually lies along the upper arc, which appears fainter in the WL map. The back projected trajectory (on a Parker spiral with V=400 km/s) of Earth is shown in blue. We thus see that the X component of the magnetic field remains negative in the HelioCast model, while it changes sign, accordingly with the data in the simulation using Multi-VP and WindPredict-AW. 

These 3D secondary structures (apart from the HCS) are due to the quasi-separatrices network \citep{PriestDemoulin1995, Demoulin1996} which can be visualized computing the squashing factor Q, shown in the bottom left panel of Figure \ref{fig:qsls}. High values of the squashing factor indicate strong connectivity gradients and, typically, pseudo-streamers lying beneath in the low corona \citep{TitovDemoulin1999, Titov2007}. Pseudo-streamers are notorious sources of slow wind and thus should be accounted for in our method, especially since it can lure our algorithm in misplacing the Heliospheric Current Sheet \citep[see, e.g.][]{Antiochos2011}. This can be a recurrent problem, especially during periods of higher solar activity, when the detection algorithm will catch more and more pseudo-streamers that appear as bright as the main streamer belt in the WL synoptic maps. Several attempts have been (and are still being) tried out to sort out this issue: 

\begin{itemize}
    \item A height-wise approach that includes WL emissions from 2.5$R_\odot$ with LASCO-C2 up to 16$R_\odot$ with LASCO-C3, so that regular streamers can be better discriminated from pseudo-streamers as they tend to have distinct radial extents. However, this method is still not robust enough and still has a low success rate, primarily because of the low signal-to-noise ratio of LASCO-C2 observations. Nonetheless, this method should become much more effective with high-sensitive WL coronagraphs such as METIS onboard \textit{Solar Orbiter}, but also \textit{Proba-3}, \textit{PUNCH} coming in the next few years. \\

    \item An integrated approach that includes pseudo-streamers in the building process of the inner boundary condition. In that sense, the solar wind speed profile may not be applied along the SMB line alone, but also along a connected network of pseudo-streamers which are extracted beforehand from the WL map (e.g. using a lower detection threshold). Whether a different velocity-distance empirical law should be used or not for pseudo-streamers still needs to be clarified. \\

    \item And lastly, a combination of the presented method with 3-D tomography reconstructions of the coronal electron density should help at improving forecast capabilities, as recent tomography techniques have recently been proven to be viable even in a time-limited operational context \citep[see e.g.][]{Bunting2022}.
\end{itemize}

\section{Conclusion}
\label{sec:ccl}

We present, evaluate and discuss a novel method, based on white-light maps obtained with the LASCO C2 coronagraph, to model and propagate the steady solar wind from the corona up to 1 au. This empirical model derives laws for the wind velocity, density, and temperature as a function of the angular distance to the estimated HCS, which is associated with the streamer maximum brightness on the WL maps. 

Based on point-by-point statistical metrics and event based analysis, we show that this model is very efficient in predicting HSSs, given the right choice for the slow wind region thickness. It outperforms more complex ab-initio models of the solar corona, such as Multi-VP and WindPredict-AW, with probability of detection above 60\% for the period considered, the first semester of 2018. 

We expect the model performance to decrease as the solar cycle rises. The main risk is the false identification of the HCS through the SMB method, which can capture quasi-separatrice layers (QSLs) instead. It is however very encouraging that the whole QSLs structure is present in the WL maps, and our model could be improved greatly by identifying and differentiating true separatrices from QSLs. 

\begin{acknowledgements}

The IRAP team acknowledges support from the French space agency (Centre National des Études Spatiales (CNES, https://cnes.fr/fr) that funds the Centre de Données de la Physique des Plasmas (CDPP, http://cdpp.eu) and the Solar-Terrestrial Observations and Modelling Service (STORMS, http://storms-service.irap.omp.eu/). The work of VR, NP, and APR was funded by the ERC SLOW\_SOURCE project (SLOW\_SOURCE—DLV-819189). The work of NF was funded by the STFC grant ST/W001071/1, P91826.
\end{acknowledgements}

\bibliography{biblio}

\end{document}